\documentclass[aip]{revtex4-1}

\usepackage[T1]{fontenc}     
\usepackage{SIunits}
\usepackage{graphicx}
\usepackage{graphicx}
\usepackage{braket}

\begin{document}

\author{Andreas W. Schell}
\email{schell@kuee.kyoto-u.ac.jp}
\affiliation{Department of Electronic Science and Engineering, Kyoto University, 615-8510 Kyoto, Japan}
\author{Toan Trong Tran}
\affiliation{School of Mathematical and Physical Sciences, University of Technology Sydney, Ultimo, New South Wales 2007, Australia}
\author{Hideaki Takashima}
\affiliation{Department of Electronic Science and Engineering, Kyoto University, 615-8510 Kyoto, Japan}
\author{Shigeki Takeuchi}
\affiliation{Department of Electronic Science and Engineering, Kyoto University, 615-8510 Kyoto, Japan}
\author{Igor Aharonovich}
\affiliation{School of Mathematical and Physical Sciences, University of Technology Sydney, Ultimo, New South Wales 2007, Australia}

\title{Non-linear excitation of quantum emitters in two-dimensional hexagonal boron nitride}

\begin{abstract}
Two-photon absorption is an important non-linear process employed for high resolution bio-imaging and non-linear 
optics. In this work we realize two-photon excitation of a quantum emitter embedded in a two-dimensional material.
We examine defects in hexagonal boron nitride and show that the emitters exhibit 
similar spectral and quantum properties under one-photon and two-photon 
excitation. Furthermore, our findings are important to deploy two-dimensional hexagonal boron nitride 
for quantum non-linear photonic applications.

\end{abstract}

\maketitle

Atomic thin, two-dimensional (2D) materials have recently emerged as promising candidates for various 
nanophotonic and optoelectronic devices, owing to their 
strong luminescence and unique electronic structure~\cite{Xia2014,Yang2015,Gutierrez2012,Gong2014,Splendiani2010,Liu2015,Fiori2014,Tonndorf2015}. 
For instance, studies of non-linear absorption or refraction in transition metal dichalcogenides 
have been promising for applications in mode locking, ultrafast photonics, and 
optical switching~\cite{Li2016,Li2015,Zhang2015,Tsai2013}. 
 
Hexagonal boron nitride (hBN) is another layered material~\cite{Caldwell2014,Song2010} that has recently been subject to an 
increased research due its ability to host room-temperature quantum emitters~\cite{Tran2016,jungwirth2016,Bourrellier2016}. While the origin 
of these emitters is still under investigation, they exhibit remarkable properties such as 
ultra-high brightness, full polarization and tunable emission~\cite{Tran2016a} making them very interesting 
for quantum sensing and optical communications.

In this work we demonstrate for the first time non-linear excitation of a single emitter in a two-dimensional material, 
namely single defects in hBN flakes via two-photon excitation. 
In two-photon absorption, in contrast to linear one-photon 
absorption, two photons of twice the one-photon wavelength get absorbed in a non-linear process 
employing a virtual state~\cite{Kaiser1961, Friedrich1982}. This process is highly important in molecular spectroscopy, 
biological microscopy and imaging as it reduces the autoflourescence of the surrounding media\cite{Denk1990}. 
Combining two-photon excitation with defects in two-dimensional materials enables atomically thin and lightweight 
bio-labels, which can be measured with a largely reduced background under excitation and detection 
in the biological window~\cite{Weissleder2001}. 
Furthermore, since the single quantum emitters are photostable, two photon absorption is a 
promising pathway to achieve wavelength multiplexing and non linear quantum photonics with two-dimensional materials.

\begin{figure}
	\centering
		\includegraphics{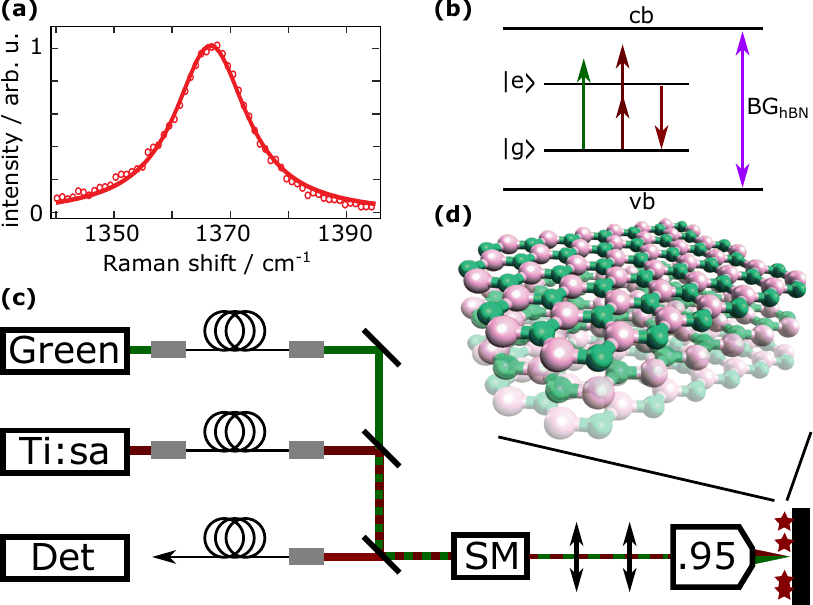}
		\caption{\textbf{hBN flakes and optical setup.}
		\textbf{a}, Raman measurement indicating the presence of multilayer hBN flakes.
		\textbf{b}, Simplified level structure of a defect in hBN. In the band gap 
		($BG_{hBN}=\unit{5.96}{\electronvolt}$) between the conduction and valence bands~\cite{Cassabois2016}  
		(cb and vb) lies a two level defect with ground and excited state ($\ket{g}$ and $\ket{e}$).
		This level can be excited by one or by two photons and will on spontaneous decay emit a photon.
		\textbf{c},
    The optical setup used consists of a home-built laser scanning confocal microscope. 
		\unit{532}{\nano\meter} wavelength cw and pulsed light can be delivered using the 
		upper beam path (Green). Also, light from a titanium sapphire laser (Ti:sa) is delivered to 
		the microscope. The excitation 
		is then coupled to a micrsocope objective with a numerical aperture of 0.95 and 
		focussed on the sample. Fluorescence light is collected with the same objective lens and 
		coupled to a fibre, which can be plugged to different detectors (Det).
		A scanning mirror (SM) in a 4f configuration is used for beam scanning.
		\textbf{d}, Structure of multilayer hBN (boron rose, nitrogen green).
		}
		\label{fig:setup}
\end{figure}

The setup used to demonstrate two-photon excitation of defects in hBN flakes consists of a home-built laser scanning confocal 
microscope with a numerical aperture of 0.95 capable of delivering laser pulses of \unit{1}{\pico\second}  duration (measured with 
an autocorrelator after fibre deliver and filtering) at 
\unit{780}{\nano\meter} wavelength with a bandwidth of \unit{2}{\nano\meter} or continuous wave light and pulses of about  \unit{50}{\pico\second} 
at \unit{532}{\nano\meter} wavelength (See Figure~\ref{fig:setup}(c)).

Fluorescence light is collected in a spectral window from \unit{561}{\nano\meter} to \unit{700}{\nano\meter} 
blocking the contributions from scattered laser light at both excitation wavelengths.

The sample investigated were prepared as follows:
A native oxide Si (100) substrate was cleaned by ultrasonication in acetone, isopropanol, and ethanol. 
Before drop-casting \unit{100}{\micro\liter} of ethanol solution containing pristine h-BN 
flakes (Graphene Supermarket) of approximately \unit{200}{\nano\meter} onto silicon substrates.
The completely dried sample was then loaded into a fused-quartz tube in a tube furnace (Lindberg Blue$\textsuperscript{TM}$). 
The tube was evacuated to low vacuum (\unit{10^{-3}}{Torr}) by means of a scroll pump then purged for \unit{30}{\minute} 
under \unit{50}{sccm} of Ar with pressure regulated at \unit{1}{Torr}. The substrate was then heated at \unit{850}{\celsius} 
for \unit{30}{\minute}  under \unit{1}{Torr} of argon. 
As last step, the sample was allowed to cool to room-temperature under continuous gas flow.
Figure~\ref{fig:setup}(a) shows a Raman spectroscopy measurement performed in order to characterize 
the sample structure. A single Lorentzian peak fitting yields \unit{1366}{cm^{-1}} suggesting the sample 
is multilayer hBN~\cite{Tran2016}. The structure of multilayer hBN is shown in Figure~\ref{fig:setup}(d)
Defects in this material are expected to have a (simplified) level structure as shown in Figure~\ref{fig:setup}(b).
The defects lie in the materials band gap of $BG_{hBN}=\unit{5.96}{\electronvolt}$~\cite{Cassabois2016} and 
can be exited in the excited state $\ket{e}$ using one or two photons before they decay to 
the ground state $\ket{g}$ under emission of a photon.

\begin{figure}
	\centering
		\includegraphics{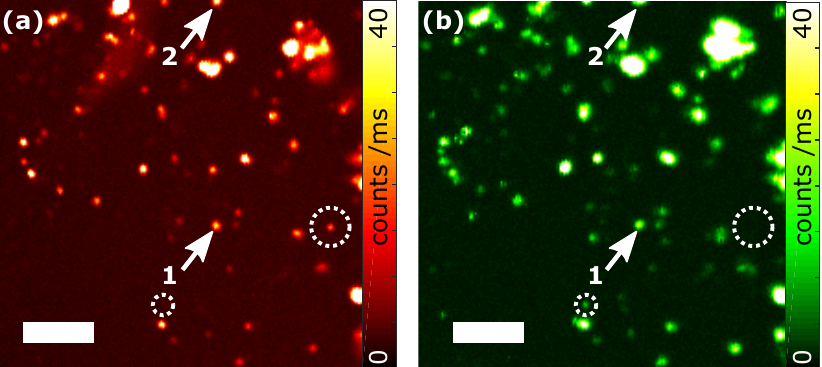}
		\caption{\textbf{Two-photon microscopy of hBN flakes.}
		\textbf{a}, Scan of a sample of hBN flakes using two-photon excitation at a wavelength of \unit{780}{\nano\meter}.
		\textbf{b}, Scan of the same area as in \textbf{a} using continuous wave excitation at \unit{532}{\nano\meter} wavelength.
		Numbered arrows indicate defects investigated in the following. Circles indicate two examples of emitter only present in one
		of the scans.
		Scale bars are \unit{5}{\micro\meter}.
		}
		\label{fig:scans}
\end{figure}

Figure~\ref{fig:scans}(a,b) shows confocal micrographs depicting the same region on a sample of hBN flakes 
excited with pulsed light at  \unit{780}{\nano\meter} and continuous wave (cw) excitation at \unit{532}{\nano\meter} 
wavelength, respectively.
Note the different relative fluorescence intensity that the same defects have in both scans. Interestingly, some of the defects 
are only visible with \unit{532}{\nano\meter} wavelength excitation and invisible under \unit{780}{\nano\meter}
excitation wavelength and vice versa.
To be able to compare the one-photon with the two-photon excitation, a defect which is clearly 
visible in both scans (indicated by a white arrow labelled 1) is investigated in the following.

\begin{figure}[t]
	\centering
		\includegraphics{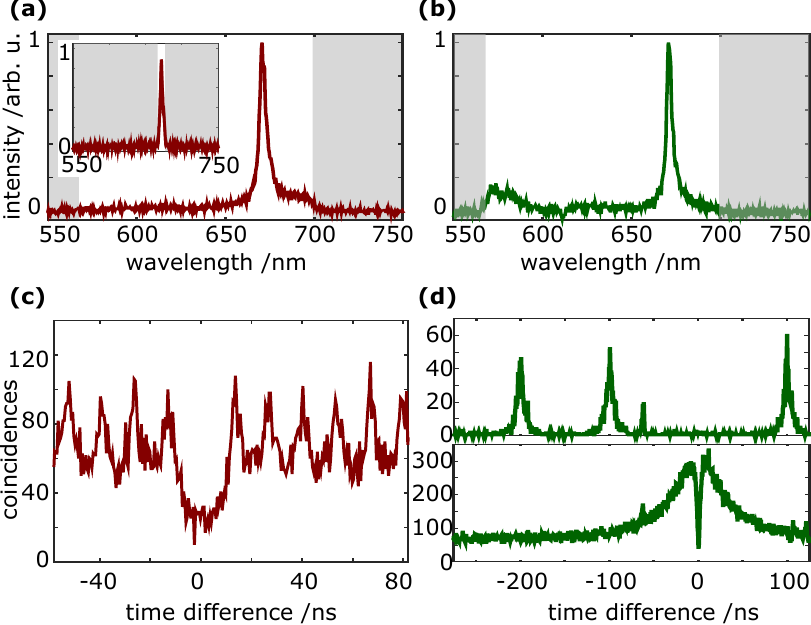}
		\caption{\textbf{Spectra and autocorrelation.}
		\textbf{a} and \textbf{b}, Spectrum of the defect labelled 1 in Figure~\ref{fig:scans}(c,d) using 
		two-photon excitation at \unit{780}{\nano\meter} and one photon excitation at \unit{532}{\nano\meter} wavelength, respectively.
		The inset in \textbf{a} shows the spectrum after a \unit{10}{\nano\meter} bandpass filter centred at the defects emission 
		wavelength had been employed. Both spectra are normalized the same way. Excitation powers were \unit{11.8}{\milli\watt} and 
		\unit{17}{\micro\watt} in \textbf{a} and \textbf{b}, respectively.
		\textbf{c}, Intensity autocorrelation measurement using pulsed red excitation.
		The absence of the peak at zero time delay clearly indicated the presence of a single photon emitter.
		\textbf{d}, Intensity autocorrelation measurement using green excitation. The upper part is
		measured using a pulsed \unit{532}{\nano\meter} wavelength laser while the lower part is measured using a 
		cw laser of the same wavelength. The small peak at 
		\unit{-60}{\nano\second} is an artefact due to cross talk of the avalanche photo diodes.
		}
		\label{fig:spectra}
\end{figure}

\begin{figure}
	\centering
		\includegraphics{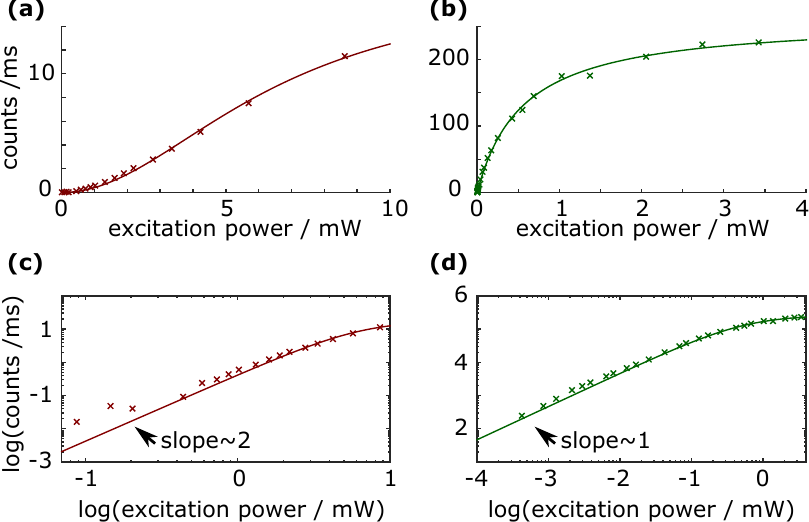}
		\caption{\textbf{Excitation power dependence of emission.}
		\textbf{a},  Dependence of the single photon emission on excitation power using \unit{780}{\nano\meter} wavelength pulsed light.
		The line is a fit to Equation~\ref{eq:tp} 
		\textbf{b}, Same as \textbf{a}, but using \unit{532}{\nano\meter} wavelength cw excitation light. The line is a fit to 
		Equation~\ref{eq:op} 
		\textbf{c} and \textbf{d}, double logarithmic plots of the data in \textbf{a} and \textbf{b}, respectively.
		Note the different sloped for low excitation powers. Integration for each data point was \unit{10}{\second}.
		}
		\label{fig:powerdependence}
\end{figure}

Figure~\ref{fig:spectra}(a) shows the spectrum of this emitter using pulsed \unit{780}{\nano\meter} wavelength 
excitation. Light in the greyed out wavelength ranges is blocked by optical filters. 
A clear peak is visible at \unit{670}{\nano\meter} (full width at half maximum approx. \unit{5}{\nano\meter}). 
The inset shows the spectrum filtered by a \unit{10}{\nano\meter} bandpass filter centred at the peak wavelength.
In Figure~\ref{fig:spectra}(b) the same emitter is shown, but using \unit{532}{\nano\meter} wavelength cw excitation.
Again, a clear peak is visible. The spectrum around the peak differs slightly between (a) and (b), what is 
attributed to excitations in the surrounding material, which are different when using different excitation wavelengths. 
The emission corresponds to an isolated defect embedded in a hBN lattice, in accordance to previous results~\cite{Tran2016a, Tran2016}. 
Figure~\ref{fig:spectra}(c) and (d) show intensity autocorrelation measurements of the photons from the defect 
with pulsed red and cw and pulsed green excitation, respectively. In these measurements, the \unit{10}{\nano\meter} bandpass filter 
was employed in order to suppress background photons.  
Excitation power was \unit{13.2}{\milli\watt} in (c), what corresponds to approximately twice the 
saturation power $I_s$ (see Figure~\ref{fig:powerdependence} and Equation~\ref{eq:tp}). 
All excitation powers are measured before the beam enters the microscope objective.
The measurements confirm that the addressed defect is a true single photon emitter. 
The complete absence of a peak in the coincidences at zero time delay for pulsed green excitation in the upper part of (d) indicates the purity of the emitter. 
From this measurement, also the excited state lifetime of the emitter is extracted to be \unit{10.4}{\nano\second}.
The cw green excitation measurement is taken in the saturated regime at an 
excitation power of \unit{3.7}{\milli\watt}.  Besides the antibunching dip (with a depth limited by our detectors' timing resolution), it shows a
pronounced bunching behaviour, which indicates the presence of a metastable state~\cite{Bernard1993,Kurtsiefer2000}. Using pulsed 
red excitation, also in the saturated regime (see Figure~\ref{fig:spectra}(c)) this behaviour is not visible, 
an indication that due to different excitation wavelengths 
different levels in the defect are addressed that are not included in the simplified 
level scheme in Figure~\ref{fig:setup}(b). One possible explanation to this is, that the shelving state present 
gets depopulated by the red laser in a kind of re-pumping process~\cite{Kurtsiefer2000}.

In order to get a better understanding of the defect and to proof that indeed two-photon excitation is the dominant 
excitation process using \unit{780}{\nano\meter} wavelength pulsed light, the dependence of the number of photons emitted 
on the excitation power was measured. 
For a one-photon process, (denoted by subscript 1) this dependence follows~\cite{Novotny2012}:
\begin{equation}
R=R_{\inf,1}\,\frac{\left(\frac{I}{I_{s,1}}\right)}{\left(\frac{I}{I_{s,1}}\right)+1}\, ,
\label{eq:op} 
\end{equation}
where $R$ is the photon emission ate, $I$ is the excitation intensity, $I_{s,1}$ is the saturation intensity, and 
$R_{\inf,1}$ is the maximum photon emission rate.
For a two-photon process (denoted by subscript 2), where the dependence on intensity is quadratic, the number of emitted photons follows:
\begin{equation}
R=R_{\inf,2}\,\frac{\left(\frac{I}{I_{s,2}}\right)^2}{\left(\frac{I}{I_{s,2}}\right)^2+1}\, .
\label{eq:tp} 
\end{equation}

The corresponding measurements are shown in Figure~\ref{fig:powerdependence}. In Figure~\ref{fig:powerdependence}(a) and (c),
the measurements using \unit{780}{\nano\meter} wavelength pulsed light are shown on a linear and double logarithmic scale, 
respectively. A fit to Equation~\ref{eq:tp} (drawn through line) yields $R_{\inf,2}=\unit{18,000}{\per\second}$ and $I_{s,2}=\unit{6.5}{\milli\watt}$. 
In Figure~\ref{fig:powerdependence}(b) and (d),
the measurements using \unit{532}{\nano\meter} wavelength cw light are shown on a linear and double logarithmic scale, 
respectively. A fit to Equation~\ref{eq:op} (drawn through line) yields $R_{\inf,1}=\unit{260,000}{\per\second}$ and $I_{s,1}=\unit{0.56}{\milli\watt}$.
All fits were performed using a least squares algorithm. 

While these measurements proof the two-photon excitation using the red laser, there is a large difference in 
the maximum count rate $R_{\inf}$ between the different modes of excitation. This difference of over one magnitude can not 
be explained by different setup configuration or due to chromatic aberrations since the emitted photons have 
the same wavelength in both cases. 
Bad alignment of the excitation hence would only influence $I_{s}$.
Therefore, as in the antibunching measurements, we assume that there are additional states present, which get populated differently 
depending on the excitation mode. Interestingly, the emitter is brighter under green excitation, 
what is the opposite one would expect if it gets de-shelved by the red laser. This is 
a indication that the defects in hBN possess a complicated level structure. Further evaluation of 
the level structure lies beyond the scope of this paper, but we want to point out that this makes 
defects in hBN highly interesting for applications in quantum optics, where it is highly desired to have the possibility to address 
different levels.

\begin{figure}
	\centering
		\includegraphics{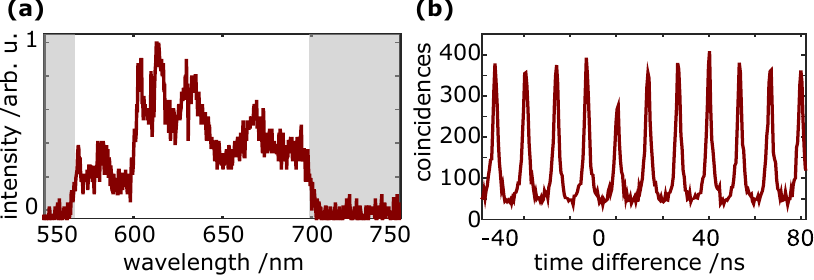}
		\caption{\textbf{Spectrum and second order autocorrelation of defect 2.}
		\textbf{a}, Spectrum of the defect labelled 2 in Figure~\ref{fig:scans} under pulsed red excitation.
		Many lines are visible on a broad background.
		\textbf{b}, Corresponding second order autocorrelation measurement. The peak at zero time 
		delay is considerably reduced proofing the non-classicality of the light emitted.
		This clearly shows that even in a small scan area as in Figure~\ref{fig:scans} 
		many quantum emitters capable of two-photon excitation can be found.
		}
		\label{fig:fourth}
\end{figure}

Figure~\ref{fig:fourth}(a) shows the spectrum of the defect labelled 2 in Figure~\ref{fig:scans} under pulsed red excitation. 
On a broad background, there are many lines visible. As expected from this, in the second order autocorrelation 
function shown in Figure~\ref{fig:fourth}(b) the peak at zero is still present, yet reduced, what means 
that more than one emitter contributes to the signal measured. This shows, that even on a 
small scan as the scan in Figure~\ref{fig:scans} many quantum emitters can be seen.
While many single emitters are observed using single photon (green) excitation,  
that is not necessarily the case for two-photon excitation. Due to the high excitation powers used for 
two-photon measurements, stability of the defects becomes an issue. While some emitters, as emitter 1, were long term stable, 
many others got destroyed by the laser. One way to solve this in future experiments would be to use a laser pulse 
shorter than the one of \unit{1}{\pico\second} used here, what will reduce thermal load on the emitters considerably. 
Nevertheless, as shown in Figure~\ref{fig:fourth}, it is possible to find more emitters that behave non-classically.

In conclusion, we have demonstrated two-photon excitation of single quantum emitters in two-dimensional 
hexagonal boron nitride. To our knowledge, this is the first time that such a measurement has 
successfully performed on a two-dimensional material or a single defect in general. 
This demonstrates the capabilities of two-dimensional materials for 
potential use as small and stable bio-markers operating in the biological window wavelength range.
Our findings achieved under different excitation conditions 
give first insights in the defects' level structure, what is a prerequisite on the 
way to non-linear quantum photonics.

\section*{Acknowledgments}
The work was supported in part by the Australian Research Council (ARC) Discovery Early Career 
Research Award (DE130100592) and the  ARC Research Hub for Integrated Device for End-user 
Analysis at Low-levels (IH150100028), FEI Company and by the AOARD grant FA2386-15-1-4044,  
MEXT/JSPS KAKENHI Grant Number 26220712, 21102007, 
Special Coordination Funds for Promoting Science and Technology, and the Cooperative 
Research Program of "Network Joint Research Center for Materials and Devices".
AWS is funded by the Japanese Society for the Promotion of Science through a fellowship 
for overseas researchers.


\end{document}